\documentclass[aps,prl,twocolumn,showpacs]{revtex4}

\usepackage{graphicx}
\usepackage[ansinew]{inputenc}

\newcommand{\buck}{C$_{60}$}

\begin{document}
\bibliographystyle{prsty}

\title{Rotation of C$_{60}$ in a Single-Molecule Contact}
\author{N. Néel}
\affiliation{Institut für Experimentelle und Angewandte Physik, Christian-Albrechts-Universität zu Kiel, D-24098 Kiel, Germany}
\author{L. Limot}
\altaffiliation{Present address: Institut de Physique et Chimie des Matériaux de Strasbourg, UMR 7504, Université Louis Pasteur, F-67034 Strasbourg, France}
\author{J. Kröger}
\email{kroeger@physik.uni-kiel.de}
\author{R. Berndt}
\affiliation{Institut für Experimentelle und Angewandte Physik, Christian-Albrechts-Universität zu Kiel, D-24098 Kiel, Germany}

\begin{abstract}
The orientation of individual C$_{60}$ molecules adsorbed on Cu(100) is
reversibly switched when the tip of a scanning tunneling
microscope is approached to contact the molecule. The probability of switching
rises sharply upon displacing the tip beyond a threshold. A mechanical
mechanism is suggested to induce the rotation of the
molecule.
\end{abstract}


\maketitle

Using single atoms or molecules as building blocks in electronic circuits
currently is of considerable interest. Experimentally, a scanning tunneling
microscope has been used to observe hopping of a Xe atom between a Ni surface
and the microscope tip \cite{dme_91}. Bistable conformational changes of
molecules have also been induced and observed with scanning tunneling
microscopy (STM) \cite{xhq_04,byc_06,mma_06,jhe_06}. Recently, the conductance
of a molecule has been controlled through the electrostatic field of a nearby
adatom \cite{pgp_05}. However, the contact regime, where the conductance approaches
the conductance quantum G$_0=2\,\text{e}^2/\text{h}$ (e: electron charge,
h: Planck's constant) has mostly been explored using mechanically controlled
break junctions \cite{mre_97,jre_02,mki_07}. A notable exception is the work
of Moresco {\it et al.}\,\cite{fmo_01}, who induced and imaged conformational
changes of a porphyrine molecule and recorded conductance data in the tunneling
and contact regimes. Overall, detailed experiments which provide information
on the geometry of single molecule switches as well as on their conductance
are scarce, in contrast to a vast body of theoretical work on molecular
conductance (see, for instance, Refs.\,\cite{yxu03a,yxu03b} and references
therein).

Here, we report on a controlled rotation of \buck\ on Cu(100) when the molecule
is brought into contact with the tip of a scanning tunneling microscope. A
sharp threshold of the tip-molecule distance above which switching occurs is
observed. It corresponds to junction conductances of
$G\approx 0.3$ -- $0.5\,\text{G}_0$. The results favor a mechanical switching
mechanism.

\begin{figure}[t]
  \includegraphics[width=85mm,clip=]{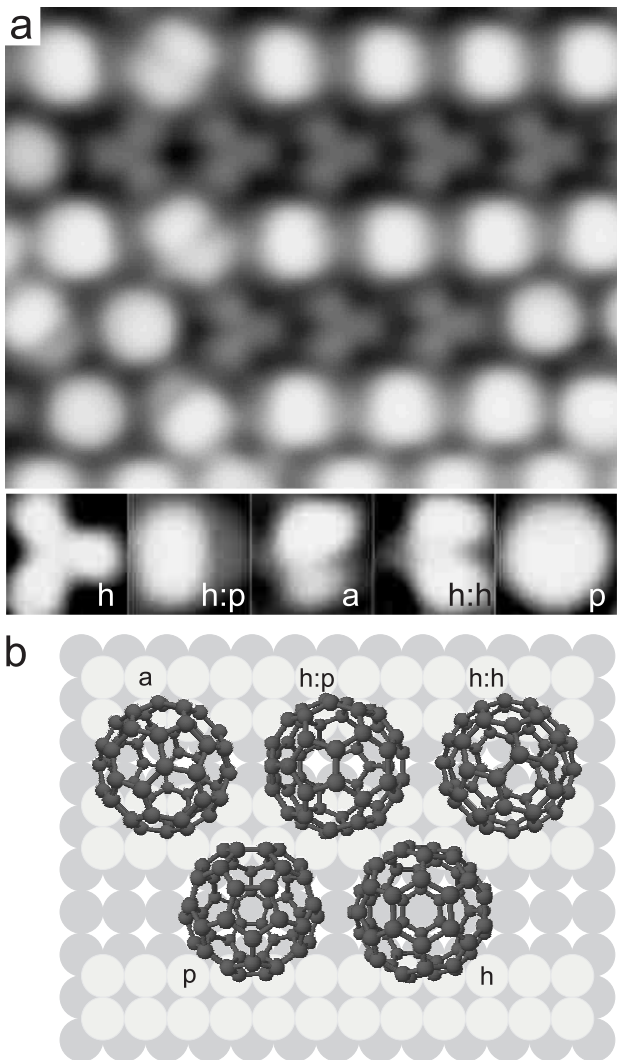}
  \caption[fig1]{(Color online) (a) STM image of \buck\ on
  Cu(100) at $T=8\,\text{K}$ after annealing at $500\,\text{K}$ (sample
  voltage $V=1.5\,\text{V}$, tunneling current $I=2.5\,\text{nA}$, size
  $60\,\text{\AA}\times 60\,\text{\AA}$). Inset: Close-up view of the five
  adsorption configurations. (b) Sketches of different \buck\ orientations
  on reconstructed Cu(100) (first (second) layer of substrate is depicted
  as bright (dark) circles). Image processing using Nanotec WSxM
  \cite{iho_07}.}
  \label{fig1}
\end{figure}

The experiments were performed using a custom-built scanning tunneling
microscope operated at $8\,\text{K}$ and in ultrahigh vacuum with a base
pressure of $10^{-9}\,\text{Pa}$. A Cu(100) surface and chemically etched
tungsten tips were cleaned by Ar ion bombardment and annealing. Tips were
further prepared {\it in vacuo} by soft indentations into the copper
surface, until intramolecular resolution of \buck\ was achieved
(Fig.\,\ref{fig1}). Given this preparation, tips were most likely
covered with substrate material. The \buck\ molecules were deposited onto
the clean surface at room temperature from a heated tantalum crucible, the
residual gas pressure remaining below $5\times 10^{-8}\,\text{Pa}$. Ordering
of \buck\ was obtained by subsequent annealing at $500\,\text{K}$.

Figure \ref{fig1}a presents constant-current STM images of \buck\ molecules
adsorbed on Cu(100). The images were acquired at $1.5\,\text{V}$ applied to
the sample. At this bias the second-to-lowest unoccupied molecular orbital
(LUMO+1) resonance is detected in spectra of the differential conductance
(not shown). The pattern therefore reflects the spatial distribution of
the density of states of this orbital \cite{jgh_99}. The molecules are
organized in a hexagonal array and form alternating bright and dim stripes
(Fig.~\ref{fig1}a). The height difference of $(0.5\pm 0.1)\,\text{\AA}$ between
the two stripes has been attributed to a missing-row reconstruction of the
copper surface following an annealing at $500\,\text{K}$ \cite{abe_03}. Bright
stripes correspond to molecules residing on a single-missing copper row, dim
stripes to molecules residing on a double-missing copper row. An inspection
of 700 molecules acquired with different tips shows that \buck\ adopts five
molecular orientations on the surface. Three additional orientations are
identified with respect to the previous study on Cu(100) \cite{abe_03}. A
close-up view is presented in the inset of Fig.\,\ref{fig1}a where distinct
structures can be seen for each orientation.
It is well established that images of the unoccupied states of \buck\ reflect
the molecular symmetry \cite{has_93,alt_93}, in particular, that bright structures
at a sample voltage close to the LUMO+1 energy are produced by the pentagon
rings \cite{jgh_99,xlu_03}. The top-most features of the STM images shown in
the inset of Fig.\,\ref{fig1}a correspond then to, from left to right, a
hexagon ring (denoted $h$), a hexagon-pentagon bond ($h$:$p$), an apex atom
($a$), a hexagon-hexagon bond ($h$:$h$), and a pentagon ring ($p$) to be compared
with the sketches in Fig.\,\ref{fig1}b. Molecules with a $h$ and $p$ orientation
are adsorbed on double-missing copper rows with a distribution of $31\,\%$ and $5\,\%$,
respectively. Molecules with $h$:$p$, $a$ and $h$:$h$ orientations are adsorbed
on single-missing copper rows with a distribution of $56\,\%$, $4\,\%$, and $4\,\%$,
respectively.

\begin{figure}[t]
  \includegraphics[width=85mm,clip=]{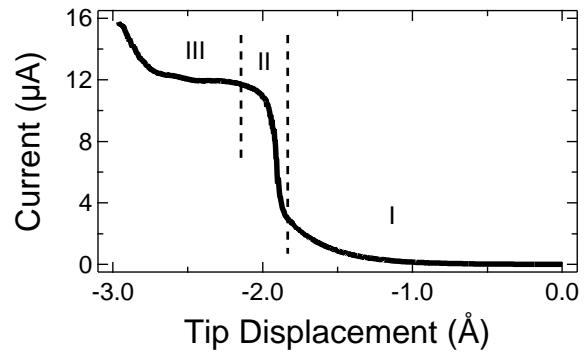}
  \caption[fig2]{Current versus tip displacement for \buck\ with $h$:$p$
  orientation. The current curve spans the tunneling (I), the transition (II),
  and the contact (III) regimes. Zero displacement corresponds to feedback
  loop parameters of $500\,\text{mV}$ and $3\,\text{nA}$.}
  \label{fig2}
\end{figure}

Having identified the molecular orientations of \buck\ on Cu(100), current
versus tip displacement measurements were performed over each orientation.
During a current measurement, the STM tip is first placed above the center
of a molecule, the feedback loop is then opened, and the tip is approached
toward the molecule at a given sample voltage simultaneously recording the
current. Figure \ref{fig2} shows a typical current curve as a function of the
tip displacement. For currents below $\approx 3\,\mu\text{A}$ the current
exhibits an exponential behavior (region I in Fig.\,\ref{fig2}). Within a
one-dimensional description of the tunneling barrier where
$I\propto\exp(-1.025\sqrt{\Phi}\Delta z)$ ($I$: current, $\Delta z$: tip
displacement) an apparent barrier height of $\Phi=(10.2\pm 0.7)\,\text{eV}$
may be extracted. Above $\approx 3\,\mu\text{A}$ a sharp increase of the
current up to $\approx 12\,\mu\text{A}$ is observed (region II), signaling
the formation of a bond between a carbon atom of \buck\ and a copper atom at
the tip apex \cite{nne_07}. Once the bond is established the contact regime
is reached (region III). The molecular contact exhibits a conductance of
$\approx 0.3\,\text{G}_0$. In the contact region we find a plateau of nearly
constant current which starts to rise above a certain tip displacement again.
The width and the slope of the plateau as well as the contact conductance
depend on the tip shape and on the location where the contact to the molecule
was formed.

\begin{figure}[t]
  \includegraphics[width=85mm,clip=]{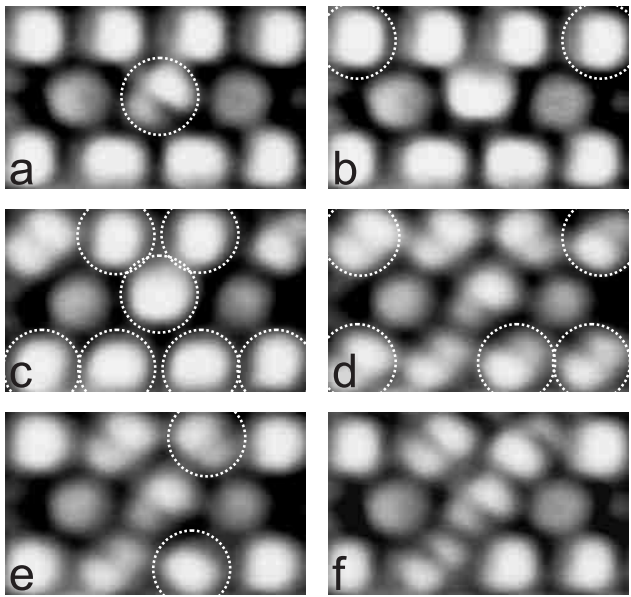}
  \caption[fig3]{(Color online) Constant-current STM images ($V=1.7\,\text{V}$,
  $I=0.1\,\text{nA}$, $35\,\text{\AA}\times 20\,\text{\AA}$) of the same area
  of the surface. (a) surface prior to switching experiments; (b)-(f) after
  contacting the molecules encircled by dashed lines in (a)-(e).}
  \label{fig3}
\end{figure}

We observed that approaching the tip by $\approx 1.9\,\text{\AA}$ beyond the
transition region (II) often causes the molecule to rotate as illustrated by
the STM images of Fig.\,\ref{fig3}. The encircled molecule in Fig.\,\ref{fig3}a
was contacted by the tip of the microscope. Increasing the tip displacement above a
threshold leads to a switch of the adsorption geometry from an $a$ to a $h$:$p$
configuration (see Fig.\,\ref{fig3}b). Contacting the two $h$:$p$ molecules
encircled in Fig.\,\ref{fig3}b leads to switching of these molecules to $h$:$h$
and $a$ configurations (see Fig.\,\ref{fig3}c). In some rare cases the molecules
rotate in the surface plane, as seen in Figs.\,\ref{fig3}e and \ref{fig3}f.
Our observations show that the switching of $a$ and $h$:$h$ molecules always
leads to $h$:$p$ molecules, while switching of $h$:$p$ molecules leads to $h$:$h$
or $a$ species, and no $h$ nor $p$ configurations have been observed. However,
contact between the tip and a molecule of the dark row ($h$ or $p$) did not
lead to a switching of the molecule adsorption configuration. Modifying the
tip apex shape by indentation of the tip into the substrate surface led to
the same observations of molecular switching, indicating that this phenomenon
is rather tip-independent.

\begin{figure}[t]
  \includegraphics[width=85mm,clip=]{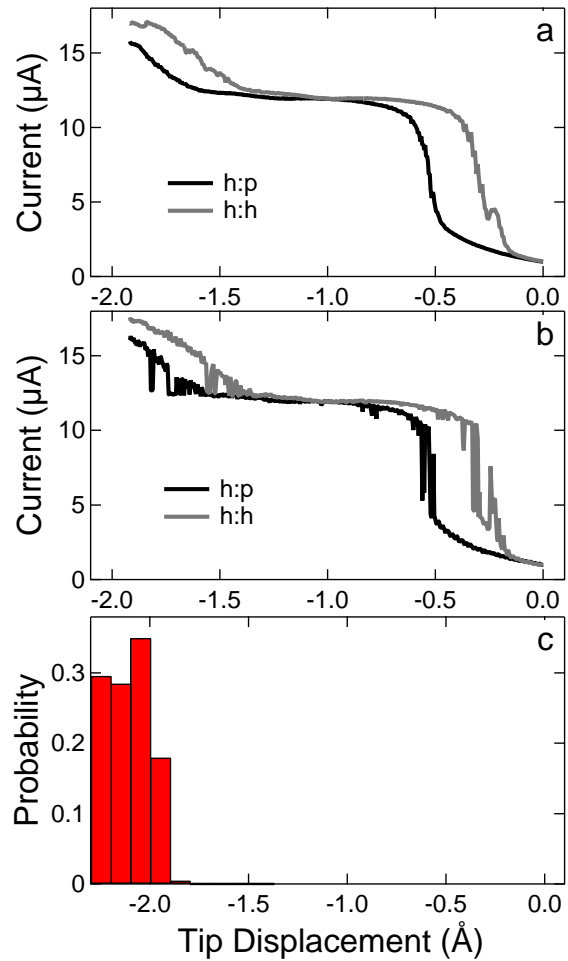}
  \caption[fig4]{(Color online) (a) Averaged current versus tip displacement
  curves acquired on top of a $h$:$p$ (black curve) and a $h$:$h$ (gray
  curve) configuration. For both curves, the feedback loop is opened at
  $300\,\text{mV}$ and $1\,\mu\text{A}$. (b) Single current versus displacement
  curve of $h$:$p$ molecule (black) together with single current curve
  characteristic of the $h$:$h$ configuration (gray). (c) Switching probability
  as a function of the tip displacement.}
  \label{fig4}
\end{figure}

The adsorption configuration of the \buck\ molecule directly determines the
current characteristics. While this observation holds for all $\text{C}_{60}$
adsorption configurations, we restrict the discussion to the $h$:$h$ and $h$:$p$
orientations below. Figure \ref{fig4}a shows averaged conductance curves
acquired on a $h$:$p$ (black curve) and on a $h$:$h$ (gray curve) molecule.
For both measurements the feedback is opened at a current of $1\,\mu\text{A}$
and a sample voltage of $300\,\text{mV}$. For the used tip the molecules exhibit
almost the same contact conductance of $\approx 0.5\,\text{G}_0$. However,
for the $h$:$h$ molecule the contact is established at a smaller tip displacement
because of its closer initial tip-molecule distance. Therefore, each adsorption
configuration is recognizable through its conductance curve, provided that
the feedback loop parameters are the same for all configurations studied.
It is then possible to observe the switching events by measuring subsequent
conductance curves on top of the selected molecule, without imaging the molecule.
This is illustrated in Fig.\,\ref{fig4}b. The current curves are assigned to
a $h$:$p$ (black) and a $h$:$h$ (gray) molecule. The black curve was acquired
just before a switching event, while the gray curve was taken directly after.
Both curves exhibit fluctuations in the transition region as well as in the
contact region when the current starts to rise again. These fluctuations are
smoothed in the averaged curves shown in Fig.\,\ref{fig4}a. Fluctuations in
the transistion region II have been interpreted in terms of local heating of
the molecule \cite{nne_07}. These fluctuations do not lead to rotation of the
molecule. The fluctuations in the contact region III, however, occur at tip
displacements around the threshold and rotation is often observed. Therefore,
these sudden changes of the current are likely to reflect switching events.

To investigate the probability of switching for different tip displacements
we used the fact that the conductance curve is characteristic of the adsorption
configuration. For each value of the displacement with same initial conditions
500 conductance measurements were performed and the number of switching events
was counted. The switching probability (Fig.\,\ref{fig4}c) rises sharply above
a threshold displacement of $\Delta z\approx -1.9\,\text{\AA}$ to $\approx 30\,\%$
rather independent of the tip displacement. Measurements with larger tip
displacements were shown to lead to damage of the tip and of the contacted
area.

Below we argue that the molecular rotation is mechanically induced whereas
local heating plays a minor role in exciting rotations. An analysis of
conductance curves on \buck\ at $300\,\text{mV}$ \cite{nne_07} showed that
energy dissipation in the tip-molecule junction leads to an effective heating
of the junction which could cause rotation of the molecule.  However, in the
present experiments we varied the total power dissipated by a factor of 40.
The switching probability was found to be insensitive to the dissipated power.
Although only a small fraction of the total power is dissipated directly at
the molecule this finding suggests that thermal excitation alone is not the
driving force for switching. Moreover, the tunneling current was not observed
to be decisive for inducing rotation. We therefore suggest that mechanical
contact with the tip causes \buck\ to rotate.

Mechanically induced rotation explains that certain rotation angles are less
frequently observed. For instance, switching a \buck\ from $h$:$p$ to $h$:$h$
or vice versa requires a rotation by 20.6°, while a smaller angle of 11.6°
is required for a rotation from $h$:$p$ to $a$ orientation. The apparent
rotation by 90° in the surface plane (Figs.\,\ref{fig3}e and \ref{fig3}f) can
also be achieved by an out-of-plane rotation of 19.2° and 36.0° for the $h$:$p$
and $h$:$h$ orientations, respectively. The large angle needed for the apparent
in plane rotation of the $h$:$h$ molecule is consistent with the observed low
frequency of this switching event. In the case of the $h$:$p$ molecule the
apparent in plane rotation results in a different adsorption geometry on the
copper surface, with now the carbon-carbon bond between the hexagon and the
pentagon of the molecule parallel to the copper missing row. This specific
orientation of the $h$:$p$ molecule is rarely observed (see Fig.\,\ref{fig1}a)
indicating a less favorable adsorption energy. This explains why this specific
event of switching is rare despite the relatively small angle needed to induce
an apparent in plane rotation of $90^{\circ}$. Switching between the $h$ and
$p$ molecules would require a relatively large rotation angle of 37.4°. In
addition, their adsorption on top of the two-missing copper rows leads to a
stronger bonding with the surface. The bonding is not limited to only the top
most copper atoms of the surface as for the three other configurations but also
occurs with the copper atoms at the bottom of the missing rows \cite{tfr_06}.
The large angle needed for the rotation and the higher coordination of these
configurations with the surface can then explain the absence of switching events
between the $h$ and $p$ configuration upon elevated tip displacements.

The current curves presented in Fig.\,\ref{fig4}a show after a relatively
flat current plateau a much faster rise of the current. In this region, in
particular for displacements larger than the threshold reproducibility of the
measurements strongly depends on the tip shape as well as on the position of
the contact over the molecule. Different slopes and shapes of this current
rise were observed. However, for all measurements this continuous rise was
observed to lead to a -- on the 100 $\mu$s time scale of data acquisition --
discontinuous jump of the current at higher tip displacements. The discontinuous
jump of the current may be attributed to a modified geometry of the contact
atomic structure \cite{nne_07}. It is reasonable to assume that during the
continuous rise of the current preceding the rearrangement of the contact
atomic structure the molecule adsorption geometry is already slightly modified.
Once the tip is retracted the molecule returns to one of its possible stable
configurations. The switching event occurs when this final configuration
differs from the initial one before the contact with the tip.

In conclusion, we observed reversible, mechanically controlled switching of
the orientation of \buck\ in a single-molecule contact. Switching occurs
when the tip of the scanning tunneling microscope is approached beyond a
well-defined threshold. This threshold is located in the first conductance
plateau where $G\approx 0.3$ -- $0.5\,\text{G}_0$.

This work was funded by Deutsche Forschungsgemeinschaft through SFB 677.

\end{document}